# Control of antiferromagnetic spin axis orientation in bilayer Fe/CuMnAs films

P. Wadley[1], K.W. Edmonds[1], M.R. Shahedkhah[1], R.P. Campion[1], B.L. Gallagher[1], J. Zelezny[2,3], J. Kunes[4,5], V. Novak[2], T. Jungwirth[2,1], V. Saidl[6,2], P. Nemec[6], F. Maccherozzi[7] and S. S. Dhesi[7]

[1]School of Physics and Astronomy, University of Nottingham, Nottingham NG7 2RD, United Kingdom

[2]Institute of Physics, Academy of Sciences of the Czech Republic, Cukrovarnická 10, 162 00 Praha 6, Czech Republic

[3] Max Planck Institute for Chemical Physics of Solids, 01187 Dresden, Germany

[4] Institute of Physics, Academy of Sciences of the Czech Republic, Na Slovance 1999/2, 182 21 Praha 8, Czech Republic

[5] Institute of Solid State Physics, TU Wien, Wiedner Hauptstr. 8, 1040 Wien, Austria

[6] Faculty of Mathematics and Physics, Charles University in Prague, Ke Karlovu 3, 121 16 Prague 2, Czech Republic

[7]Diamond Light Source, Chilton, Didcot, Oxfordshire, OX11 0DE, United Kingdom

*Using x-ray magnetic circular and linear dichroism techniques, we demonstrate a collinear exchange coupling between an epitaxial antiferromagnet, tetragonal CuMnAs, and an Fe surface layer. A small uncompensated Mn magnetic moment is observed which is antiparallel to the Fe magnetization. The staggered magnetization of the 5nm thick CuMnAs layer is rotatable under small magnetic fields, due to the interlayer exchange coupling. This allows us to obtain the x-ray magnetic linear dichroism spectra for different crystalline orientations of CuMnAs in the (001) plane.*

Antiferromagnetic (AF) spintronics is an emerging field which aims to utilize the particular properties of AF materials for information storage and processing applications [1]. The collinear antiferromagnet tetragonal CuMnAs is of particular interest due to its crystal structure, in which the two Mn spin sublattices form inversion partners in a centrosymmetric lattice (Fig. 1(a)) [2]. Due to spin-orbit coupling, an electric current results in a local spin polarization, of opposite sign on each sublattice, which can induce a torque large enough to rotate the staggered magnetization between stable configurations [3,4,5]. Further, theoretical studies have predicted the presence of Dirac band crossings in both the tetragonal and orthorhombic phases of CuMnAs, co-existing with and influenced by the AF order [6,7]. Methods to control the AF order in CuMnAs are therefore or substantial current interest.

The physics of exchange coupling at a ferromagnetic (FM) / antiferromagnetic (AF) interface has been widely studied, both for fundamental understanding and for applications in magnetic storage and memory technologies. Characteristic features of such interfaces include enhancement of the coercivity and a shift of the hysteresis loop (exchange bias) of the FM layer [8]. Studies of epitaxial interfaces between crystalline materials offer particular insights, due to their well-controlled interface structures and magnetocrystalline anisotropies [9]. The configuration of the spins in the AF layer – whether bulk or surface, fully anti-aligned or partially uncompensated, rotatable or frozen in place – can strongly affect the behavior of the FM layer [10]. The AF configuration in FM/AF bilayers has been explored directly, using x-ray spectroscopy and spectromicroscopy techniques as well as tunnelling anisotropic magnetoresistance [11,12,13,14]. Such experiments have shown the close connection between the rotation and pinning of AF moments and the hysteresis of the FM layer. This also provides a means to manipulate the spins in an AF layer for potential spintronic applications [9,15].

Ab initio calculations indicate that the stable configurations of the staggered magnetization in tetragonal CuMnAs lie in the (001) plane, where a biaxial magnetic anisotropy is expected due to the crystal symmetry [2,16]. However, the tetragonal polytype of CuMnAs is stabilized by growth on III-V substrates (GaP or GaAs), which leads to an in-plane uniaxial magnetic anisotropy [16,17]. Similar anisotropies are commonly found in FM/III-V films, due to the broken symmetry of the III-V surface [18].

Here we present a study of the interlayer exchange coupling in a bilayer film consisting of FM Fe and AF CuMnAs. We combine x-ray magnetic circular dichroism (XMCD) and x-ray magnetic linear dichroism (XMLD) to obtain element specific information on the FM layer as well as both compensated and uncompensated magnetic moments in the AF layer. In crystalline materials, the XMLD in particular contains rich information on the atomic and magnetic structure. Crystalline anisotropy of XMLD spectra, in which the spectral lineshape depends strongly on the direction of the x-ray polarization vector with respect to the crystallographic axes, has been observed in theoretical and experimental studies of a wide variety of magnetic materials including metals [19,20], oxides [21,22,[23]] and diluted magnetic semiconductors [24]. Here we utilize the exchange coupling between the Fe layer and rotatable AF CuMnAs spins to reveal the anisotropic XMLD spectra for tetragonal CuMnAs, which are compared to ab initio calculations.

The sample studied consists of a 2 nm Al / 2 nm Fe / 5 nm CuMnAs film grown on a GaP(001) substrate by molecular beam epitaxy. The substrate temperature during growth was 260°C for the CuMnAs layer and 0°C for the Fe layer and the protective Al cap. The layers were grown in the same ultra-high vacuum chamber, to ensure a clean interface between them. Previous studies have shown that tetragonal CuMnAs is lattice-matched to GaP(001) through a 45° rotation of the unit cell [2]. The measurements described below confirm the epitaxial relationship Fe(001)[110] || CuMnAs(001)[100] || GaP(001)[110]. Figure 1(b) shows magnetization loops for the film measured by superconducting quantum interference device (SQUID) magnetometry along one of the in-plane <110> directions of the GaP substrate. Negligible exchange bias is observed, while the coercivity is around 50 Oe at 200K and 70 Oe at 2K. The rounded shape of the loop is ascribed to crystalline disorder, due to the large lattice mismatch between Fe and GaP(001).

The XMCD and XMLD measurements were performed on beamline I06 of Diamond Light Source, using total electron yield detection and a superconducting vector magnet in which magnetic fields can be applied in any direction. XMCD spectra were measured with the x-ray beam at a grazing angle of 25° to the sample surface, and with a magnetic field of 1000 Oe applied along the beam direction, as illustrated in Fig. 1(c). Figures 1(d) and 1(e) show the Mn $L_{2,3}$ and Fe $L_{2,3}$ x-ray absorption and XMCD spectra from the sample, at a sample temperature of 250K. The Mn XMCD is very weak and of opposite sign to the Fe XMCD, indicating a small net Mn magnetic moment which is antiferromagnetically coupled to the Fe layer. The antiparallel alignments of the Fe and CuMnAs magnetic moments is in contrast to Fe$_{1-x}$Mn$_x$ binary alloys, for which the Mn moment is small and parallel to the Fe [25]. The magnitude of the XMCD asymmetry $(I^+ - I^-)/(I^+ + I^-)$, where $I^+$ and $I^-$ are the Mn $L_3$ peak heights above background for photon helicity parallel and antiparallel to the magnetic field, is around 1%.

As shown in Fig. 1(a), the magnetic structure in CuMnAs consists of FM (001) planes which are AF coupled to the neighboring sublattice planes. Therefore, the interface plane of CuMnAs may be expected to consist of uncompensated Mn magnetic moments. Due to the finite probing depth of the total electron yield XMCD measurement, the signal from the uncompensated interface layer is not fully cancelled by the opposite oriented layer below it. The XMCD from the AF ordered CuMnAs film will be smaller than for a fully FM oriented CuMnAs film by a factor $R = (1 - e^{-a/d})/(1 + e^{-a/d})$, where $a$ is the sublattice plane spacing and $d$ is the total electron yield probing depth. Taking $a$ = 0.3 nm and $d \approx$ 3nm [26] gives $R \approx 0.05$, consistent with the small size of the observed Mn XMCD. However, we do not rule out a possible contribution from rotatable uncompensated moments in the bulk of the AF layer, or interfacial alloying.

The XMLD spectra were obtained with the x-ray beam at normal incidence, taking the difference between absorption spectra measured with the x-ray linear polarization vector parallel to the [110] and [1-10] axes of the GaP substrate. A 1000 Oe magnetic field was applied along either the [110] or [1-10] axes, with a small out-of-plane tilt in order to increase the electron yield signal. It was verified that the small out-of-plane component of the field did not affect the spectra. The experimental geometry is illustrated in Fig. 2(a). The XMLD spectra at the Mn $L_{2,3}$ and Fe $L_{2,3}$ edges at 250K are shown in Figs. 2(b) and 2(c) respectively. The XMLD spectra are shown as a fraction of the $L_3$ absorption peak height above background.

The Mn $L_3$ XMLD signal is larger than that of the Fe and comparable to that of a 10 nm CuMnAs single layer [16]. Given the large size of the Mn XMLD signal, it can be inferred that it is due to the compensated antiferromagnetic Mn moments in the CuMnAs film rather than the small number of uncompensated moments at the interface. Most strikingly, the same XMLD signal, but with opposite sign, is observed when the applied magnetic field is applied in the orthogonal direction. The reversal of the XMLD spectrum is expected for the FM layer if the Fe magnetization orients parallel to the magnetic field. The observation of similar behavior for the Mn XMLD indicates that the staggered magnetic moments in the CuMnAs layer have a uniaxial orientation and are exchange coupled to the Fe layer, following the reorientation of the Fe magnetization under the applied magnetic field. The rotation of the AF spins is also observed at 300K and 2K, although the magnitude of the XMLD is slightly reduced compared to its value at 250K, as shown in the inset to Fig. 2(c). The smaller value at 2K may be due to competition between the interlayer exchange coupling and magnetocrystalline anisotropy in the CuMnAs layer.

Figure 3 compares XMLD spectra for x-ray polarization and applied magnetic fields along the GaP substrate [100] and [110] crystal axes. For both the Mn and Fe $L_3$ absorption edges, the sign and lineshape of the XMLD depends strongly on the crystallographic direction. The Fe $L_{2,3}$ XMLD spectra shown in Fig. 3(a) are in good agreement with previous studies of epitaxial Fe films on GaAs(001) [20]. This confirms that the Fe layer is epitaxial with in-plane crystal directions parallel to those of the substrate.

The Mn $L_{2,3}$ XMLD spectra shown in Fig. 3(b) are compared to *ab initio* calculations shown in Fig. 3(c). The theoretical XMLD spectra were obtained from LDA+U electronic structure calculations [2] using the approach of Ref. 13, which neglects electronic correlations and core hole effects. The finite core hole lifetime was mimicked by lorentzian broadening of 0.4 eV. The calculations reproduce some of the main features in the experimental spectra, including the relative separations of the main peaks and their reversal in sign between the different crystal orientations. Additional features on the low-energy side of the $L_2$ and $L_3$ peaks in the calculated spectra are not observed in the experiment. The experimental XMLD spectra are defined as the absorption spectra for parallel x-ray polarization and applied magnetic field, minus the absorption spectra for perpendicular x-ray polarization and applied magnetic field. Similarly, the calculated XMLD are the absorption for AF moments parallel to x-ray polarization, minus the absorption for AF moments perpendicular to polarization. Taking into account the 45° rotation of the CuMnAs crystal with respect to the GaP substrate [2], the sign of the main peaks is in agreement between theory and experiment for both crystal orientations. The comparison of the measured spectra to the calculation therefore indicates that the AF spin axis in the CuMnAs layer is aligned collinear with the external magnetic field, *i.e.*, the interlayer exchange coupling favors a collinear alignment of the FM Fe and AF CuMnAs magnetic moments.

From the XMCD and XMLD results described above, we can infer the following. The measured XMCD is consistent with the interface atomic layer of the CuMnAs film orienting antiparallel to the epitaxial FM Fe layer as well as to the neighboring CuMnAs magnetic sublattice. The AF CuMnAs spins have a collinear coupling to the Fe layer. The AF spins in the CuMnAs layer are rotatable by reorienting the Fe magnetization under relatively small external magnetic fields. This is in contrast to for example CoO/Fe epitaxial layers, where the AF spin configuration is largely frozen for thicknesses above ≈ 3 nm [13]. Interlayer exchange coupling therefore provides a means to rotate the orientation of compensated AF materials, which are hard to manipulate directly using external magnetic fields. For tetragonal CuMnAs, this may be combined with manipulation of the magnetic order using spin-orbit torques [9,27], and electrical [9] or magneto-optical [11] detection, for future hybrid FM/AF spintronic applications.

We acknowledge Diamond Light Source for provision of beamtime under proposal SI9993 and support from the Ministry Education of the Czech Republic Grant No. LM2015087 and from the Grant Agency of the Czech Republic Grant No. 14-37427. PW acknowledges support from the University of Nottingham Engineering and Physical Sciences Research Council Impact Acceleration account (Grant EP/K503800/1).

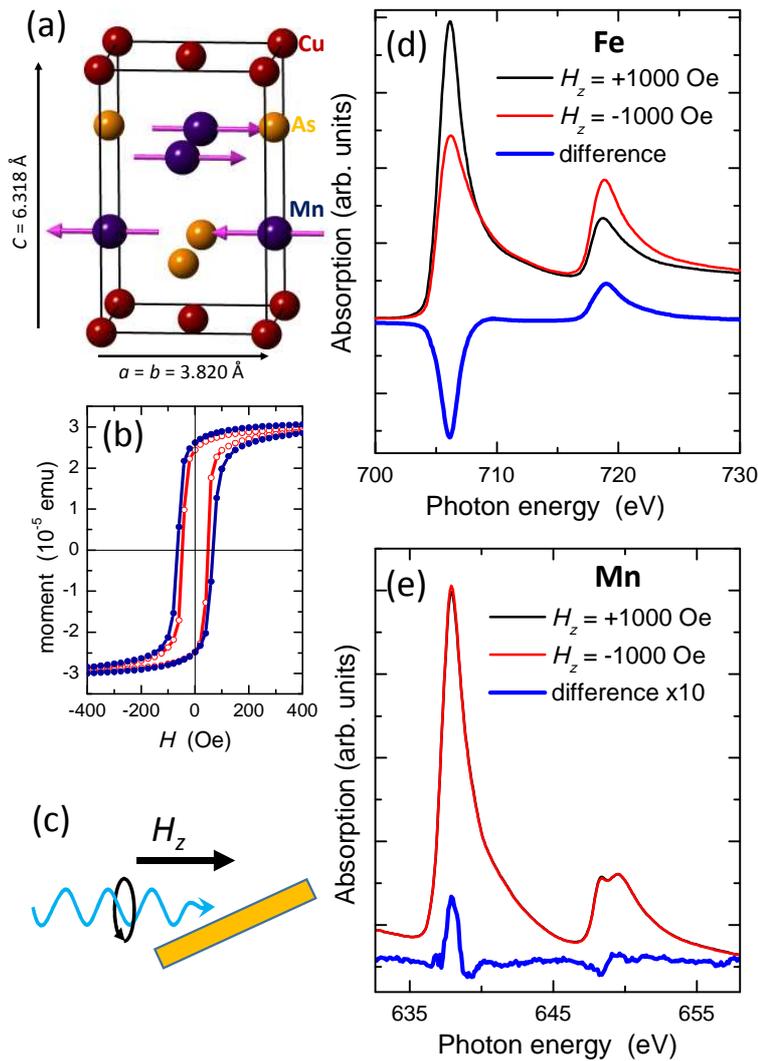

Fig. 1. (Color online) (a) Crystal structure of tetragonal CuMnAs. (b) SQUID hysteresis loops for magnetic field in the plane of the Fe/CuMnAs film, at temperature 200K (open circles) and 2K (filled circles). (c) Experimental geometry for the XMCD measurements. (d) Fe $L_{2,3}$ and (e) Mn $L_{2,3}$ absorption spectra for magnetic fields applied parallel and antiparallel to the x-ray helicity vector, and the difference (XMCD) spectra, at sample temperature 250K. The Mn XMCD is scaled by a factor of 10 for clarity.

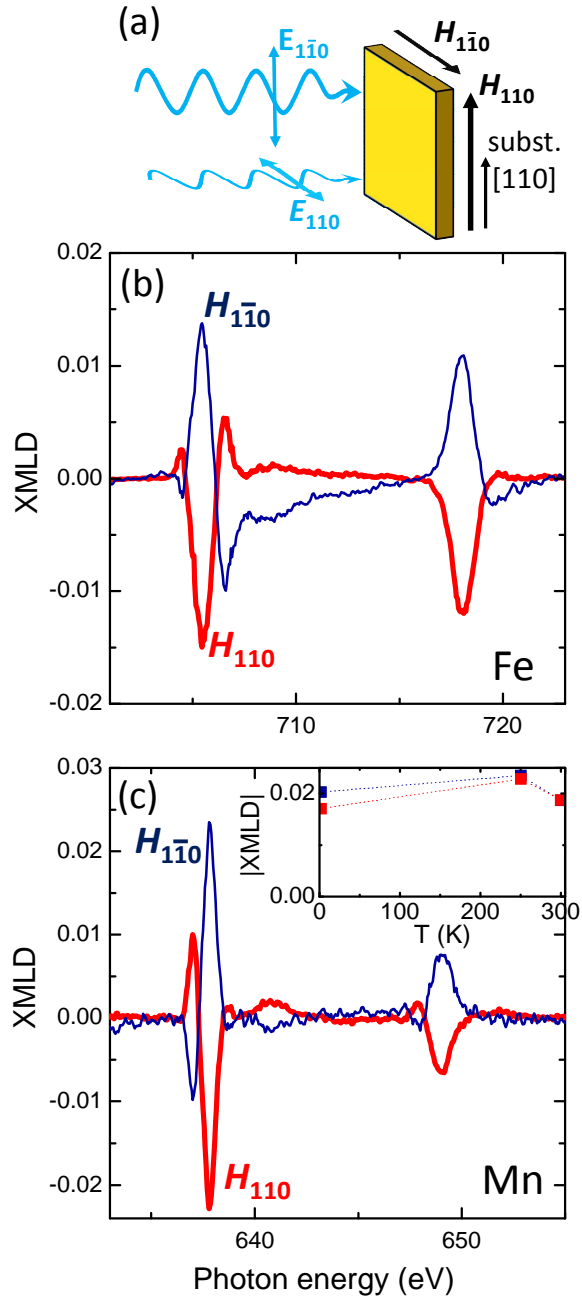

Fig. 2. (Color online) (a) Experimental geometry for the XMLD measurements. The XMLD spectra were obtained as the difference between absorption spectra with x-ray linear polarization vector $E_x$ and $E_y$. (b) Fe $L_{2,3}$ and (c) Mn $L_{2,3}$ XMLD spectra, with applied magnetic field along $x$ (thin lines) and along $y$ (thick lines), at a sample temperature is 250K. The inset to (c) shows the magnitude of the Mn $L_3$ XMLD peak as a function of temperature.

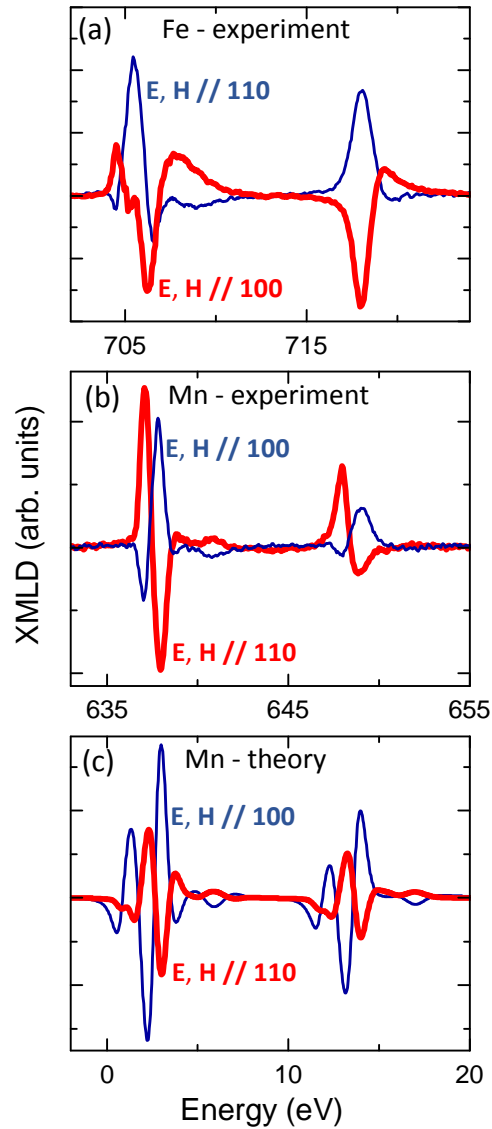

Fig. 3. (Color online) Anisotropic XMLD spectra: (a) measured Fe $L_{2,3}$ XMLD; (b) measured Mn $L_{2,3}$ XMLD; (c) calculated Mn $L_{2,3}$ XMLD. The experimental XMLD are measured at temperature $T$ = 250K, and are the difference in absorption between parallel and perpendicular configurations of the x-ray polarization and the 1000 Oe applied magnetic field, for fields along the [110] (thin blue lines) and [100] (thick red lines) in-plane orientations of the GaP substrate. Note that the CuMnAs unit cell is rotated 45° compared to the substrate. The calculated XMLD are the difference in absorption between parallel and perpendicular configurations of the x-ray polarization and the antiferromagnetic spin axis, for [100] (thin blue line) and [110] (thick red line) orientations of a tetragonal CuMnAs crystal.